\documentclass[aps,prb,twocolumn,superscriptaddress]{revtex4-2}
\usepackage{graphicx}  
\usepackage[inkscapearea=page, inkscapelatex=false]{svg}
\usepackage{relsize}
\usepackage{minitoc} 
\noptcrule 

\usepackage[export]{adjustbox}
\usepackage[caption=false]{subfig}
\usepackage{hyperref}  
\hypersetup{colorlinks=true, citecolor=blue, urlcolor=blue, linkcolor=blue}
\usepackage{dcolumn}   
\usepackage{bm}        
\usepackage{amssymb}   
\usepackage{amsmath}   
\usepackage{euscript}
\usepackage{verbatim}
\usepackage{setspace}
\usepackage{xcolor}
\usepackage{amsfonts}
\usepackage{braket}
\DeclareUnicodeCharacter{2212}{-}
\newcommand{\STO}{SrTiO$_3$}
\newcommand{\LAO}{LaAlO$_3$}

\newcommand{\pcco}{Pr$_{1.85}$Ce$_{0.15}$CuO$_{4-\delta}$}
\newcommand{\PCCO}{Pr$_{1.85}$Ce$_{0.15}$CuO$_{4-\delta}$}
\newcommand{\SAO}{Sr$_3$Al$_2$O$_6$}

\newcommand{\etal}{\textit{et al.}}
\allowdisplaybreaks

\begin{abstract}
    
Thin films of cuprate superconductors are easier to control in terms of doping as compared to bulk samples. However, they require specific substrates to facilitate epitaxial growth. These substrates are often incompatible with materials used in electronic applications. Furthermore, it is challenging to separate the substrate's properties from the material of interest. Here, we demonstrate the fabrication of an electron-doped cuprate membrane. We show that the membrane has a coherent crystal structure. Furthermore, the superconducting properties of the membrane post-liftoff closely resemble those of the thin films pre-lift-off, as revealed by a scanning superconducting quantum interference device (SQUID) microscope. Such membranes pave the way for designing new material properties and incorporating complex superconducting materials into typically incompatible electronic devices.
\end{abstract}
\begin{document}

\title{Freestanding single-crystal superconducting electron-doped cuprate membrane}

\author{S. Sandik\textcolor{red}{$^*$}}
\affiliation{%
  School of Physics and Astronomy, Tel Aviv University, Tel Aviv, 6997801, Israel\\
 }%
\author{Bat-Chen Elshalem\textcolor{red}{$^*$}}
\affiliation{%
  Department of Physics and Institute of Nanotechnology and Advanced Materials, Bar-Ilan University, Ramat-Gan 5290002, Israel\\
 }
 \author{A. Azulay}
\affiliation{%
  Department of Materials Science and Engineering, The Iby and Aladar Fleischman Faculty of Engineering, Tel Aviv University, Tel Aviv 6997801, Israel\\
 }%
\author{M. Waisbort}
\affiliation{%
  School of Physics and Astronomy, Tel Aviv University, Tel Aviv, 6997801, Israel\\
 }%

 \author{A. Kohn}
\affiliation{%
  Department of Materials Science and Engineering, The Iby and Aladar Fleischman Faculty of Engineering, Tel Aviv University, Tel Aviv 6997801, Israel\\
 }%
 \author{B. Kalisky}
\email{Corresponding author: beena@biu.ac.il}
\affiliation{%
  Department of Physics and Institute of Nanotechnology and Advanced Materials, Bar-Ilan University, Ramat-Gan 5290002, Israel\\
 }
\author{Y. Dagan}
\email{Corresponding author: yodagan@tauex.tau.ac.il}
\affiliation{%
  School of Physics and Astronomy, Tel Aviv University, Tel Aviv, 6997801, Israel\\
 }

 \date{\today}
\def\thefootnote{*}\footnotetext{\textcolor{red}{These authors contributed equally to this work}}

\maketitle

\begin{figure*}
		\includegraphics[width=1\textwidth]{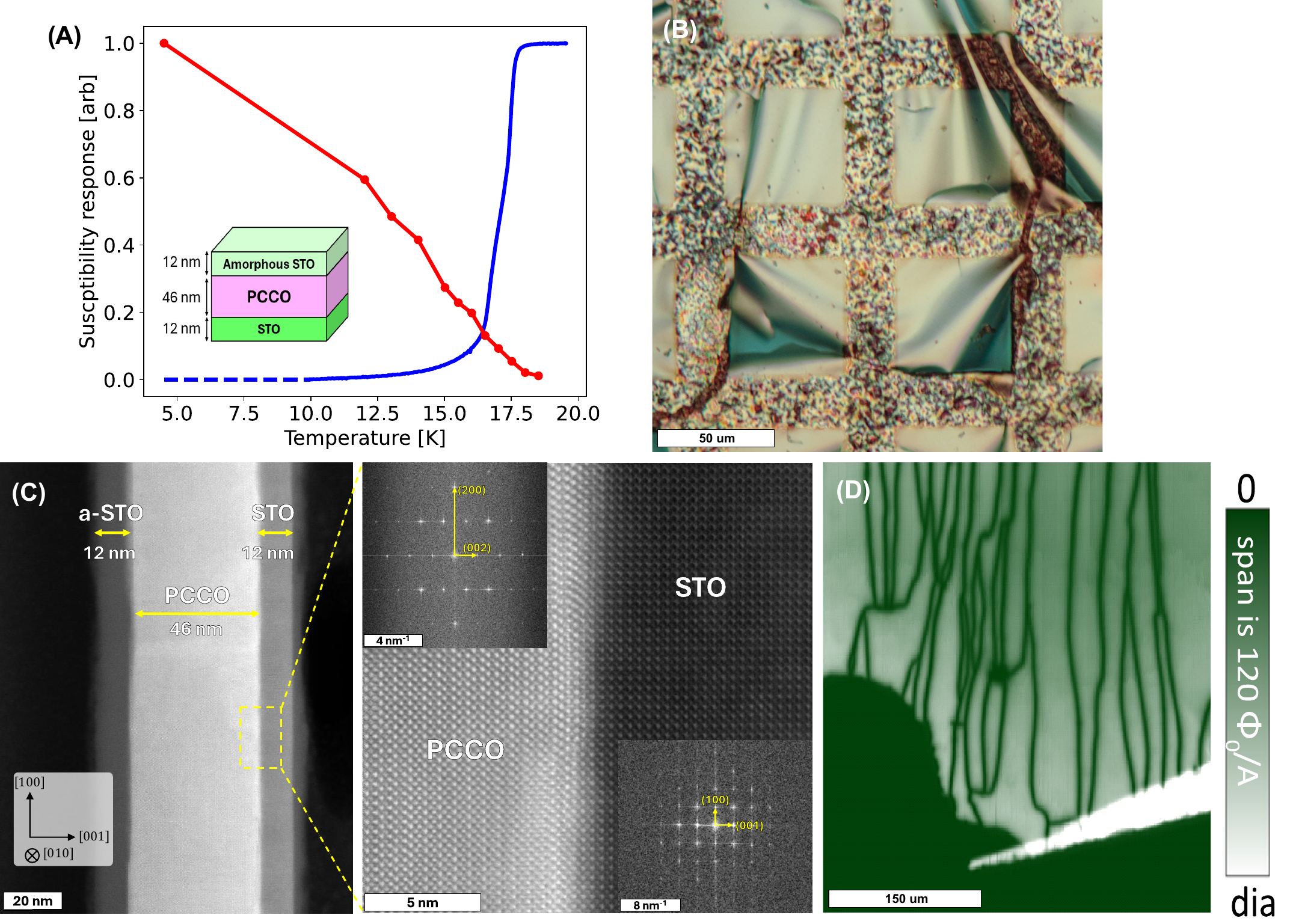}
	\caption{\pcco{} membrane. (a) Mutual induction measurement (solid blue. extrapolated data in dashed line) before releasing the membrane T$_c$ of 16.5-17K is observed. Similar T$_c$ is observed in the SQUID susceptibility (red, locally measured at position 1 marked in Fig. \ref{SQUIDSUS}a). \textbf{inset:} sketch of the structure of the membrane. (b) Optical image of a \pcco{} on a support grid. (c) A cross-sectional HAADF Z-contrast STEM image of the released membrane aligned to the [010] zone axis. The higher magnification, atomic-column resolution image to the right is from the region denoted by the yellow dashed rectangle. \textbf{inset:} Power spectra from regions in the image of the membrane and the \STO{} underlayer demonstrating an epitaxial relation of \pcco{}(001)[100]//\STO{}(001)[100]. (d) Scanning SQUID  susceptometry scan at 4.3K showing a diamagnetic response (lighter regions). The black veins are cracks in the membrane. The brighter area is a membrane folded on itself. }
	\label{allresults}
 
\end{figure*}

In the cuprates, the parent compound is an antiferromagnetic insulator. It becomes conducting and superconducting when adding charge carriers (doping) into the CuO$_2$ planes.  These charge carriers can be either holes, as for La$_{2-x}$Ba$_{x}$CuO$_4$\cite{bednorz1986possible} or electrons as for Pr$_{2-x}$Ce$_{x}$CuO$_{4-\delta}$ \cite{tokura1989superconducting,armitage2010progress}. In both cases, superconductivity depends non-monotonically on the carrier density, creating a superconducting dome in the carrier-density versus temperature phase diagram. For electron-doped cuprates, it is easier to control the doping level in the form of thin films compared to the bulk material \cite{dagan2004evidence,jin2011link}. However, it is difficult to separate the properties of the films from their bulk substrate. 

Di Lu \etal \cite{lu2016synthesis} show that a water soluble sacrificial layer of \SAO{} can be used to grow free-standing epitaxial oxide layers. Other methods were developed for separating oxide films from their substrates \cite{bakaul2016single,pesquera2020large,bourlier2020transfer,Chiabrera2022review}. These technologies sparked research of free-standing oxides such as free-standing conducting \LAO/\STO{} interfaces \cite{sambri2020self} structural changes and enhanced polarization in ultra-thin BiFeO$_3$ \cite{ji2019freestanding}, room temperature ferroelectricity in ultra-thin \STO{} films \cite{lee2015emergence}, super-elasticity in BaTiO$_3$ \cite{dong2019super}, tunable magnetism in extremely strained manganites \cite{hong2020extreme} and superconducting lanthanum-doped \STO{} membrane \cite{li2021stabilization}. Recently, a free-standing hole-doped cuprate YBa$_2$Cu$_3$O$_{7-\delta}$ heterostructure has been realized \cite{PhysRevMaterials.3.060801}. While these cuprates have a higher T$_c$ compared to the electron-doped ones, studying free-standing membranes of both electron- and hole-doped cuprates will provide us with a more complete picture of the cuprate phase diagram, thus leading to a better understanding of high-temperature superconductivity. Furthermore, an important advantage of the electron-doped cuprates is their accessible upper critical field, which makes it easier to study their normal state properties, including those hidden under the superconducting dome.
\par
One of the significant difficulties with free-standing membranes is determining whether the membrane retains its properties after releasing from the supporting substrate. The physical properties of the membrane may change due to the release \cite{hong2017two} and sudden change in strain and thickness. It is also challenging to measure such membranes due to their small signal relative to a bulk crystal and the difficulty of putting contacts that can withstand strain. 

In this letter, We demonstrate the preparation and transfer of a 46nm thick electron-doped cuprate \pcco{} membrane. Structural characterization by a Scanning Transmission Electron Microscope (STEM) reveals a coherent crystalline structure with a well-controlled composition. We utilize a scanning SQUID microscope to approach the membrane noninvasively and measure its local properties. We find that the membrane is  homogeneous with a robust $d-wave$ type superconductivity, which is typical of these types of materials \cite{tsuei2000phase}.

In Figure \ref{allresults}, we summarize our main findings. The superconducting transition temperature, T$_c$, of the heterostructure before releasing the membrane as measured by a mutual induction probe as well as SQUID susceptometry verifying that T$_c$ remains unaltered after releasing and placing the membrane onto a silicon wafer, as shown in Figure \ref{allresults} (a) (see also Figure \ref{SQUIDSUS} (b)). The inset depicts the compositing layers of the membrane and their thickness. An optical image of a \pcco{} membrane on top of a support grid is shown in Figure \ref{allresults} (b). A High-Angle Annular Dark-Field (HAADF) Z-contrast STEM image of a cross-section of the released membrane is shown in Figure \ref{allresults} (c) with a higher magnification image, to the right, showing atomic-column resolution, as well as from the bottom \STO{} layer (see inset of power spectra). These results demonstrate the high crystalline quality of the membrane and the epitaxial relation to the underlying \STO{} layer.  In Figure \ref{allresults} (d), we show scanning SQUID susceptometry (see methods for details) where the diamagnetic response of the superconducting membrane is shown. This demonstrates robust and homogeneous superconductivity.

\section{Results and discussion}

	
\begin{figure}
		\includegraphics[width=1\columnwidth]{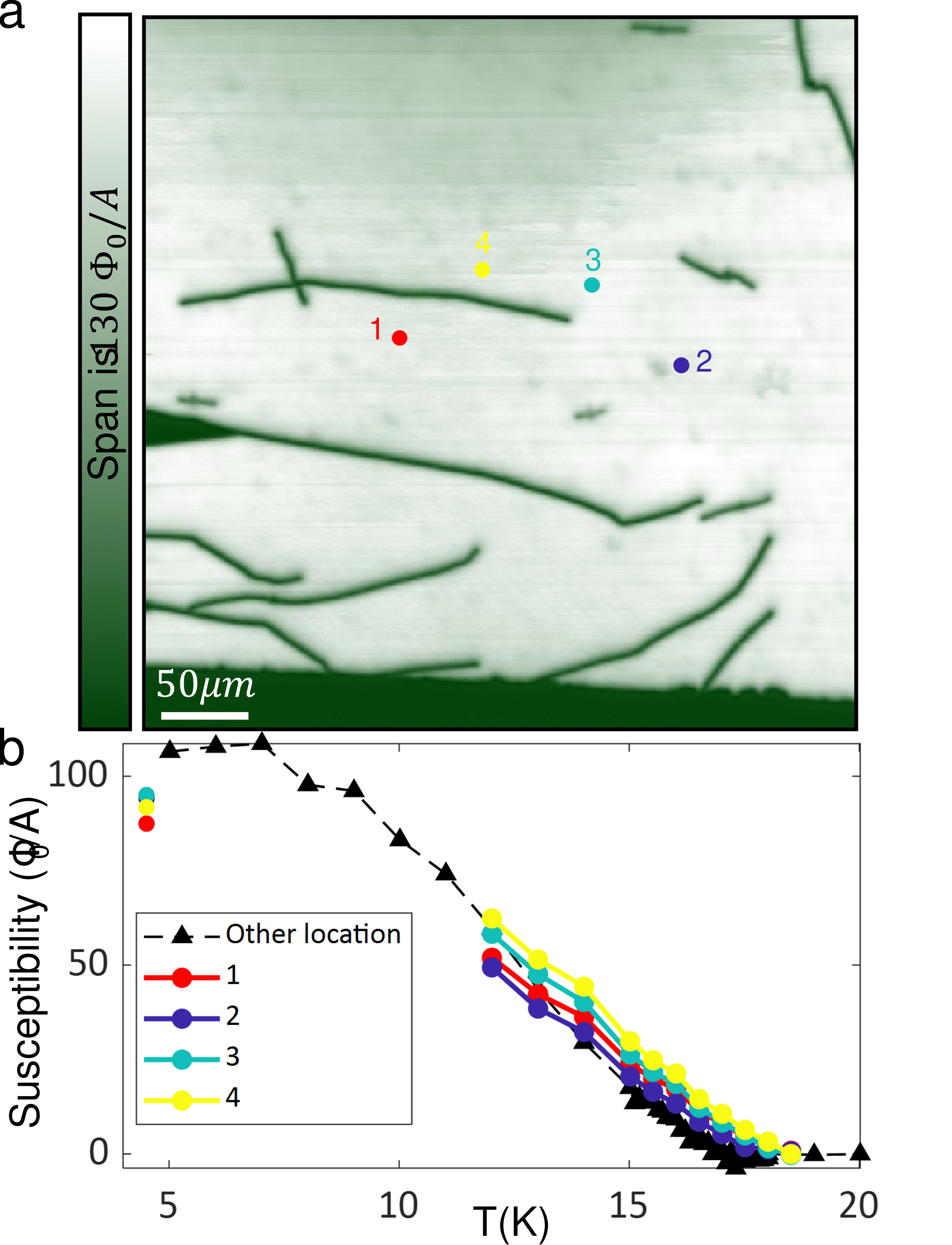}
	\caption{
The temperature dependence of the diamagnetic response. (a) Scanning SQUID map of the diamagnetic response of a $400\times 400 \left(\mu m\right)^2$ region of the membrane. Dark color at the bottom of the scan and in the cracks is zero susceptibility signal. (b) The diamagnetic reading was extracted from scans of the same location at different temperatures, shown here for four locations marked in a. Pos 1 is the same data as in FIG \ref{allresults} A. The diamagnetic response is proportional to the superfluid density. The dashed line is taken from a different location on the sample at a wider temperature range.}
	\label{SQUIDSUS}
\end{figure}

\begin{figure}
		\includegraphics[width=1.05\columnwidth]{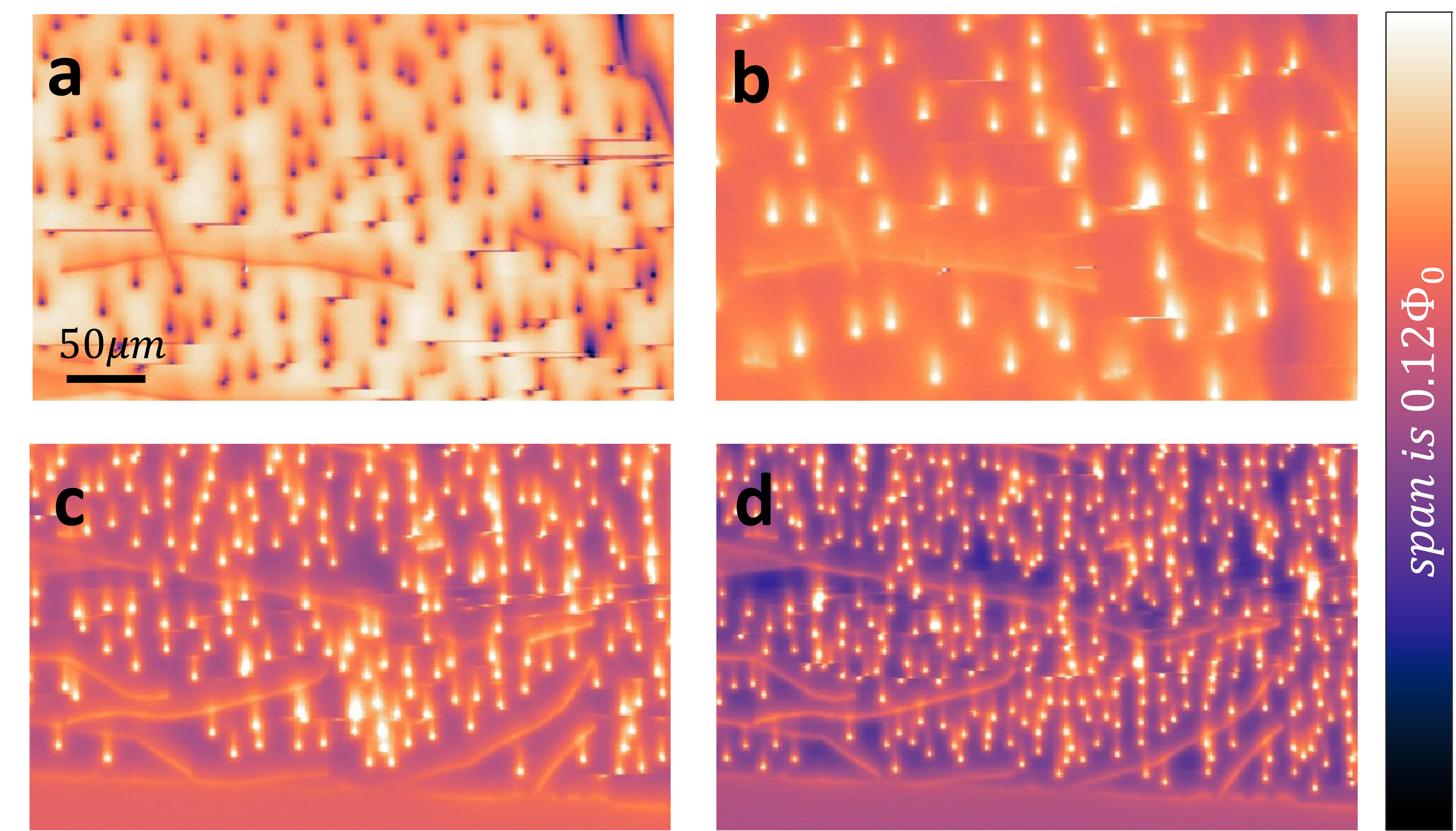}
	\caption{Vortex pin on the strongly superconducting parts. (a-d) Scanning SQUID magnetometry scan at 4K showing Meissner screening at edge and membrane cracks, and Abrikosov vortices pinned in random locations. The sample was cooled down through Tc at various fields (-32.8, 12.1, 67.3 and 119 mG for a-d, accordingly) to achieve different vortex densities.}
	\label{SQUIDVORTEX}
	
\end{figure}

\par

Following deposition and preceding the lift-off, the \pcco{} heterostructure and its supporting substrate was positioned in a mutual inductance probe consisting of two coils in order to ascertain the superconducting transition temperature T$_c$. In Figure \ref{allresults}(a), we present the measured AC susceptibility response, as measured by this setup. A distinct drop in the real component of the mutual inductance signal is observed, pointing to a macroscopic and robust superconducting transition.

After lift-off \STO/\pcco/amorphous-\STO{} flakes were transferred onto a silicon wafer. Figure \ref{allresults}(b) shows an optical image of such flakes transferred onto a TEM grid. This demonstrates the ability to transfer large-scale free-standing membranes onto devices of interest.

Scanning SQUID susceptometry of the membrane showed a clear diamagnetic signal (see Figure \ref{allresults}(d)). This signal was relatively uniform across the sample but interrupted by channels where the membrane is cracked, and no diamagnetic susceptibility is observed (the dark green lines in Figure \ref{allresults}(d). 

We followed the evolution of the diamagnetic response with temperature and observed a gradual decrease of the signal until its disappearance at T$_c$ which was set at 16.5 K (Figure \ref{SQUIDSUS}). A d-wave symmetry was shown for this material \cite{tsuei2000phase}. The temperature dependence of the diamagnetic signal further supports this observation.

In some areas of the sample, where the membrane has been folded on itself, a higher critical temperature of 17 K was demonstrated as can be seen in Figure \ref{allresults}(d) and the lower right corner of Figure \textcolor{blue}{S3} in the supplementary.

 Mapping the static magnetic flux at the same area of the susceptibility map in (Figure \ref{SQUIDSUS}a), we followed the spatial configuration of vortices, which form in the superconductor after cooling it in the presence of magnetic field (Figure \ref{SQUIDVORTEX}). Scans (a-d) at different vortex densities all show that these vortices pin in the membrane itself, and absent from the cracks. These images also show that the cracks behave like an edge, letting normal flux in.

Figure \ref{allresults}(c) (left) displays a cross-sectional HAADF Z-contrast STEM image, showing an overview of the sample depicted schematically in the inset of Figure \ref{allresults}(a) and aligned to the [010] zone axis. Figure \ref{allresults}(c) (right) provides a smaller field of view from the interface region denoted by the dashed yellow rectangle, at high spatial resolution, which demonstrates the high crystalline quality of the layers. Figure \ref{allresults}(c) insets depict power spectra calculated from this \pcco{} and \STO{} region. The location and symmetry of the frequencies of the power spectra allows to identify the tetragonal I4/mmm \pcco{} \cite{uzumaki1991crystal} and cubic $Pm\bar{3}m$ \STO{} \cite{lytle1964x} crystals.
The highlighted frequencies correspond to interplanar spacings of $d_{200}$ $\sim$ 630 and $d_{002}$ $\sim$ 210, and $d_{100}$ $\sim$ 410 pm, which are attributed to their reciprocal lattice vectors $g_{200}$ and $g_{002}$, and $g_{100}$ and $g_{001}$, respectively. Thus, we conclude that the epitaxial relation between the two layers is \pcco{}(001)[100]//\STO{}(001)[100]. 
\section {Experimental Section}
\subsection{Sample growth and transfer.}
The samples were prepared using the Pulsed Laser Deposition (PLD) technique on (001)-oriented atomically flat TiO$_2$-terminated \STO{} substrates.  The thickness of \SAO{} was calibrated using Reflection High Energy Electron Diffraction (RHEED) oscillations, as demonstrated in supplementary Figure S1. Following the \SAO{} deposition, \STO{} was deposited as a capping layer and a barrier between the \SAO{} and the subsequent \pcco{} films as explained below. 
The sample was then transferred in ambient atmosphere to a second chamber for \pcco{} deposition and annealing to remove apical oxygen \cite{higgins2006role}.
For more details, see Supplementary information. 

As demonstrated in \cite{LAdiffusion}, the thermally induced diffusion of La and Mn atoms from the thin film studied to the sacrificial layer creates a structural change, rendering the sacrificial layer insoluble. Given this observation and recognizing that water negatively affects the properties of cuprate thin films  \cite{YBCOWater}, we determined that a more effective approach is to apply thin 10nm \STO{} buffer layers on both sides of the \pcco{} thin film. This modification aims to expedite the lift-off process and preserve the film's properties even after immersion in water.

Spin-coated PMMA and PDMS were used for the lift-off process. Subsequently, the \SAO{} was dissolved in de-ionized water, and the membrane was transferred onto a Si wafer. The PDMS was then peeled off, and the PMMA layer was dissolved using acetone.

\subsection{SQUID and TEM characterisation} 
We utilized a scanning SQUID susceptometer to spatially map the magnetism and susceptibility of the \PCCO{} membrane. The SQUID gradiometer employs a pickup loop to convert magnetic flux, measured in units of $\Phi_0 =hc/(2e)$, into a detectable electric signal. We determined the local magnetic field by maneuvering the pickup loop across the sample and recording the magnetic flux at each pixel. Simultaneously, we measured the diamagnetic response. To achieve this, we employed a concentric coil surrounding the pickup loop, applying an alternating current that generated an oscillating magnetic field of a fraction of a Gauss. The pickup loop recorded the diamagnetic response from the sample.

Structural characterization and compositional analysis were carried out using an aberration-corrected scanning transmission electron microscope (STEM). The cross-sectional sample of the membrane for the STEM measurements was prepared by a Ga$^{+}$ focused ion beam.

\section{Summary}

We demonstrate that membranes of the electron-doped cuprate, \pcco{}, can be fabricated and transferred to the desired device or substrate. These membranes are single-crystalline and exhibit superconductivity, with properties closely resembling those of the thin films prior to membrane release. We further illustrate that scanning SQUID serves as a potent tool for the contactless characterization of miniaturized superconducting devices.

\begin{acknowledgments}
S.S, M.W, and Y.D are supported by the Israel Science Foundation (ISF) under grants \# ISF-1711/23 and \# ISF-476/22 and by the Oren family Chair for experimental physics.
B.E and B.K. were supported by the European Research Council \# ERC-2019-OG-866236, the ISF, \# ISF-228/22, German Israeli Project Cooperation DIP \#KA 3970/1-1, and the COST Action CA21144.
A. K. acknowledges support by the ISF under grant \# ISF-2973/21.
\end{acknowledgments}

\bibliography{main} 

\begin{thebibliography}{10}
\expandafter\ifx\csname url\endcsname\relax
  \def\url#1{\texttt{#1}}\fi
\expandafter\ifx\csname urlprefix\endcsname\relax\def\urlprefix{URL }\fi
\providecommand{\bibinfo}[2]{#2}
\providecommand{\eprint}[2][]{\url{#2}}

\bibitem{bednorz1986possible}
\bibinfo{author}{Bednorz, J.~G.} \& \bibinfo{author}{M{\"u}ller, K.~A.}
\newblock \bibinfo{title}{Possible high t c superconductivity in the ba- la- cu- o system}.
\newblock \emph{\bibinfo{journal}{Zeitschrift f{\"u}r Physik B Condensed Matter}} \textbf{\bibinfo{volume}{64}}, \bibinfo{pages}{189--193} (\bibinfo{year}{1986}).

\bibitem{tokura1989superconducting}
\bibinfo{author}{Tokura, Y.}, \bibinfo{author}{Takagi, H.} \& \bibinfo{author}{Uchida, S.}
\newblock \bibinfo{title}{A superconducting copper oxide compound with electrons as the charge carriers}.
\newblock \emph{\bibinfo{journal}{Nature}} \textbf{\bibinfo{volume}{337}}, \bibinfo{pages}{345--347} (\bibinfo{year}{1989}).

\bibitem{armitage2010progress}
\bibinfo{author}{Armitage, N.}, \bibinfo{author}{Fournier, P.} \& \bibinfo{author}{Greene, R.}
\newblock \bibinfo{title}{Progress and perspectives on electron-doped cuprates}.
\newblock \emph{\bibinfo{journal}{Reviews of Modern Physics}} \textbf{\bibinfo{volume}{82}}, \bibinfo{pages}{2421} (\bibinfo{year}{2010}).

\bibitem{dagan2004evidence}
\bibinfo{author}{Dagan, Y.}, \bibinfo{author}{Qazilbash, M.}, \bibinfo{author}{Hill, C.}, \bibinfo{author}{Kulkarni, V.} \& \bibinfo{author}{Greene, R.}
\newblock \bibinfo{title}{Evidence for a quantum phase transition in pr 2- x ce x cuo 4- $\delta$ from transport measurements}.
\newblock \emph{\bibinfo{journal}{Physical Review Letters}} \textbf{\bibinfo{volume}{92}}, \bibinfo{pages}{167001} (\bibinfo{year}{2004}).

\bibitem{jin2011link}
\bibinfo{author}{Jin, K.}, \bibinfo{author}{Butch, N.}, \bibinfo{author}{Kirshenbaum, K.}, \bibinfo{author}{Paglione, J.} \& \bibinfo{author}{Greene, R.}
\newblock \bibinfo{title}{Link between spin fluctuations and electron pairing in copper oxide superconductors}.
\newblock \emph{\bibinfo{journal}{Nature}} \textbf{\bibinfo{volume}{476}}, \bibinfo{pages}{73--75} (\bibinfo{year}{2011}).

\bibitem{lu2016synthesis}
\bibinfo{author}{Lu, D.} \emph{et~al.}
\newblock \bibinfo{title}{Synthesis of freestanding single-crystal perovskite films and heterostructures by etching of sacrificial water-soluble layers}.
\newblock \emph{\bibinfo{journal}{Nature materials}} \textbf{\bibinfo{volume}{15}}, \bibinfo{pages}{1255--1260} (\bibinfo{year}{2016}).

\bibitem{bakaul2016single}
\bibinfo{author}{Bakaul, S.~R.} \emph{et~al.}
\newblock \bibinfo{title}{Single crystal functional oxides on silicon}.
\newblock \emph{\bibinfo{journal}{Nature communications}} \textbf{\bibinfo{volume}{7}}, \bibinfo{pages}{10547} (\bibinfo{year}{2016}).

\bibitem{pesquera2020large}
\bibinfo{author}{Pesquera, D.} \emph{et~al.}
\newblock \bibinfo{title}{Large magnetoelectric coupling in multiferroic oxide heterostructures assembled via epitaxial lift-off}.
\newblock \emph{\bibinfo{journal}{Nature Communications}} \textbf{\bibinfo{volume}{11}}, \bibinfo{pages}{3190} (\bibinfo{year}{2020}).

\bibitem{bourlier2020transfer}
\bibinfo{author}{Bourlier, Y.} \emph{et~al.}
\newblock \bibinfo{title}{Transfer of epitaxial srtio3 nanothick layers using water-soluble sacrificial perovskite oxides}.
\newblock \emph{\bibinfo{journal}{ACS applied materials \& interfaces}} \textbf{\bibinfo{volume}{12}}, \bibinfo{pages}{8466--8474} (\bibinfo{year}{2020}).

\bibitem{Chiabrera2022review}
\bibinfo{author}{Chiabrera, F.~M.} \emph{et~al.}
\newblock \bibinfo{title}{Freestanding perovskite oxide films: Synthesis, challenges, and properties}.
\newblock \emph{\bibinfo{journal}{Annalen der Physik}} \textbf{\bibinfo{volume}{534}}, \bibinfo{pages}{2200084} (\bibinfo{year}{2022}).
\newblock \urlprefix\url{https://onlinelibrary.wiley.com/doi/abs/10.1002/andp.202200084}.
\newblock \eprint{https://onlinelibrary.wiley.com/doi/pdf/10.1002/andp.202200084}.

\bibitem{sambri2020self}
\bibinfo{author}{Sambri, A.} \emph{et~al.}
\newblock \bibinfo{title}{Self-formed, conducting laalo3/srtio3 micro-membranes}.
\newblock \emph{\bibinfo{journal}{Advanced Functional Materials}} \textbf{\bibinfo{volume}{30}}, \bibinfo{pages}{1909964} (\bibinfo{year}{2020}).

\bibitem{ji2019freestanding}
\bibinfo{author}{Ji, D.} \emph{et~al.}
\newblock \bibinfo{title}{Freestanding crystalline oxide perovskites down to the monolayer limit}.
\newblock \emph{\bibinfo{journal}{Nature}} \textbf{\bibinfo{volume}{570}}, \bibinfo{pages}{87--90} (\bibinfo{year}{2019}).

\bibitem{lee2015emergence}
\bibinfo{author}{Lee, D.} \emph{et~al.}
\newblock \bibinfo{title}{Emergence of room-temperature ferroelectricity at reduced dimensions}.
\newblock \emph{\bibinfo{journal}{Science}} \textbf{\bibinfo{volume}{349}}, \bibinfo{pages}{1314--1317} (\bibinfo{year}{2015}).

\bibitem{dong2019super}
\bibinfo{author}{Dong, G.} \emph{et~al.}
\newblock \bibinfo{title}{Super-elastic ferroelectric single-crystal membrane with continuous electric dipole rotation}.
\newblock \emph{\bibinfo{journal}{Science}} \textbf{\bibinfo{volume}{366}}, \bibinfo{pages}{475--479} (\bibinfo{year}{2019}).

\bibitem{hong2020extreme}
\bibinfo{author}{Hong, S.~S.} \emph{et~al.}
\newblock \bibinfo{title}{Extreme tensile strain states in la0. 7ca0. 3mno3 membranes}.
\newblock \emph{\bibinfo{journal}{Science}} \textbf{\bibinfo{volume}{368}}, \bibinfo{pages}{71--76} (\bibinfo{year}{2020}).

\bibitem{li2021stabilization}
\bibinfo{author}{Li, D.} \emph{et~al.}
\newblock \bibinfo{title}{Stabilization of sr3al2o6 growth templates for ex situ synthesis of freestanding crystalline oxide membranes}.
\newblock \emph{\bibinfo{journal}{Nano Letters}} \textbf{\bibinfo{volume}{21}}, \bibinfo{pages}{4454--4460} (\bibinfo{year}{2021}).

\bibitem{PhysRevMaterials.3.060801}
\bibinfo{author}{Chen, Z.} \emph{et~al.}
\newblock \bibinfo{title}{Freestanding crystalline $\mathrm{YB}{\mathrm{a}}_{2}\mathrm{C}{\mathrm{u}}_{3}{\mathrm{o}}_{7\ensuremath{-}x}$ heterostructure membranes}.
\newblock \emph{\bibinfo{journal}{Phys. Rev. Mater.}} \textbf{\bibinfo{volume}{3}}, \bibinfo{pages}{060801} (\bibinfo{year}{2019}).
\newblock \urlprefix\url{https://link.aps.org/doi/10.1103/PhysRevMaterials.3.060801}.

\bibitem{hong2017two}
\bibinfo{author}{Hong, S.~S.} \emph{et~al.}
\newblock \bibinfo{title}{Two-dimensional limit of crystalline order in perovskite membrane films}.
\newblock \emph{\bibinfo{journal}{Science Advances}} \textbf{\bibinfo{volume}{3}}, \bibinfo{pages}{eaao5173} (\bibinfo{year}{2017}).

\bibitem{tsuei2000phase}
\bibinfo{author}{Tsuei, C.} \& \bibinfo{author}{Kirtley, J.}
\newblock \bibinfo{title}{Phase-sensitive evidence for d-wave pairing symmetry in electron-doped cuprate superconductors}.
\newblock \emph{\bibinfo{journal}{Physical Review Letters}} \textbf{\bibinfo{volume}{85}}, \bibinfo{pages}{182} (\bibinfo{year}{2000}).

\bibitem{uzumaki1991crystal}
\bibinfo{author}{Uzumaki, T.}, \bibinfo{author}{Kamehara, N. K.~N.} \& \bibinfo{author}{Niwa, K. N.~K.}
\newblock \bibinfo{title}{Crystal structure and madelung potential in r2-xcexcuo4-$\delta$ (r= pr, nd, sm, eu and gd) system}.
\newblock \emph{\bibinfo{journal}{Japanese journal of applied physics}} \textbf{\bibinfo{volume}{30}}, \bibinfo{pages}{L981} (\bibinfo{year}{1991}).

\bibitem{lytle1964x}
\bibinfo{author}{Lytle, F.~W.}
\newblock \bibinfo{title}{X-ray diffractometry of low-temperature phase transformations in strontium titanate}.
\newblock \emph{\bibinfo{journal}{Journal of Applied Physics}} \textbf{\bibinfo{volume}{35}}, \bibinfo{pages}{2212--2215} (\bibinfo{year}{1964}).

\bibitem{higgins2006role}
\bibinfo{author}{Higgins, J.}, \bibinfo{author}{Dagan, Y.}, \bibinfo{author}{Barr, M.}, \bibinfo{author}{Weaver, B.} \& \bibinfo{author}{Greene, R.}
\newblock \bibinfo{title}{Role of oxygen in the electron-doped superconducting cuprates}.
\newblock \emph{\bibinfo{journal}{Physical Review B—Condensed Matter and Materials Physics}} \textbf{\bibinfo{volume}{73}}, \bibinfo{pages}{104510} (\bibinfo{year}{2006}).

\bibitem{LAdiffusion}
\bibinfo{author}{Baek, D.~J.}, \bibinfo{author}{Lu, D.}, \bibinfo{author}{Hikita, Y.}, \bibinfo{author}{Hwang, H.~Y.} \& \bibinfo{author}{Kourkoutis, L.~F.}
\newblock \bibinfo{title}{Ultrathin epitaxial barrier layer to avoid thermally induced phase transformation in oxide heterostructures}.
\newblock \emph{\bibinfo{journal}{ACS Applied Materials \& Interfaces}} \textbf{\bibinfo{volume}{9}}, \bibinfo{pages}{54--59} (\bibinfo{year}{2017}).

\bibitem{YBCOWater}
\bibinfo{author}{Bansal, N.~P.} \& \bibinfo{author}{Sandkuhl, A.~L.}
\newblock \bibinfo{title}{{Chemical durability of high‐temperature superconductor YBa2Cu3O7−x in aqueous environments}}.
\newblock \emph{\bibinfo{journal}{Applied Physics Letters}} \textbf{\bibinfo{volume}{52}}, \bibinfo{pages}{323--325} (\bibinfo{year}{1988}).
\newblock \urlprefix\url{https://doi.org/10.1063/1.99481}.
\newblock \eprint{https://pubs.aip.org/aip/apl/article-pdf/52/4/323/7766453/323\_1\_online.pdf}.

\end{thebibliography}

\end{document}